\documentclass[twocolumn,preprintnumbers,amsmath,amssymb,floatfix]{revtex4}

\usepackage{graphicx}% Include figure files
\usepackage{dcolumn}% Align table columns on decimal point
\usepackage{bm}% bold math

\begin{document}

\title{Cluster properties from two-particle angular correlations in p+p collisions at \boldmath $\sqrt{s} = 200~\mathrm{and}~410~\mathrm{GeV}$}

\author{
% Authors for data EXCLUSIVELY from RUN2005 which included 
% CuCu @ 22, 63 & 200 GeV and p+p @ 400 GeV
%
% Last edited 29-Sep-2005 by George Stephans\\  \vspace{0.2in}
%
B.Alver$^4$,
B.B.Back$^1$,
M.D.Baker$^2$,
M.Ballintijn$^4$,
D.S.Barton$^2$,
R.R.Betts$^6$,
R.Bindel$^7$,
W.Busza$^4$,
Z.Chai$^2$,
V.Chetluru$^6$,
E.Garc\'{\i}a$^6$,
T.Gburek$^3$,
K.Gulbrandsen$^4$,
J.Hamblen$^8$,
I.Harnarine$^6$,
C.Henderson$^4$,
D.J.Hofman$^6$,
R.S.Hollis$^6$,
R.Ho\l y\'{n}ski$^3$,
B.Holzman$^2$,
A.Iordanova$^6$,
J.L.Kane$^4$,
P.Kulinich$^4$,
C.M.Kuo$^5$,
W.Li$^4$,
W.T.Lin$^5$,
C.Loizides$^4$,
S.Manly$^8$,
A.C.Mignerey$^7$,
R.Nouicer$^2$,
A.Olszewski$^3$,
R.Pak$^2$,
C.Reed$^4$,
E.Richardson$^7$,
C.Roland$^4$,
G.Roland$^4$,
J.Sagerer$^6$,
I.Sedykh$^2$,
C.E.Smith$^6$,
M.A.Stankiewicz$^2$,
P.Steinberg$^2$,
G.S.F.Stephans$^4$,
A.Sukhanov$^2$,
A.Szostak$^2$,
M.B.Tonjes$^7$,
A.Trzupek$^3$,
G.J.van~Nieuwenhuizen$^4$,
S.S.Vaurynovich$^4$,
R.Verdier$^4$,
G.I.Veres$^4$,
P.Walters$^8$,
E.Wenger$^4$,
D.Willhelm$^7$,
F.L.H.Wolfs$^8$,
B.Wosiek$^3$,
K.Wo\'{z}niak$^3$,
S.Wyngaardt$^2$,
B.Wys\l ouch$^4$\\
\vspace{3mm}
\small
%
% Note that this is the full form of the addresses, for conference proceedings,
% you can use the reduced one that follows
%
% $^1$~Physics Division, Argonne National Laboratory, Argonne, IL 60439-4843,
% USA\\
% $^2$~Chemistry and C-A Departments, Brookhaven National Laboratory, Upton, NY
% 11973-5000, USA\\
% $^3$~Institute of Nuclear Physics, Krak\'{o}w, Poland\\
% $^4$~Laboratory for Nuclear Science, Massachusetts Institute of Technology,
% Cambridge, MA 02139-4307, USA\\
% $^5$~Department of Physics, National Central University, Chung-Li, Taiwan\\
% $^6$~Department of Physics, University of Illinois at Chicago, Chicago, IL
% 60607-7059, USA\\
% $^7$~Department of Chemistry, University of Maryland, College Park, MD 20742,
% USA\\
% $^8$~Department of Physics and Astronomy, University of Rochester, Rochester,
% NY 14627, USA\\
%
%
$^1$~Argonne National Laboratory, Argonne, IL 60439-4843, USA\\
$^2$~Brookhaven National Laboratory, Upton, NY 11973-5000, USA\\
$^3$~Institute of Nuclear Physics PAN, Krak\'{o}w, Poland\\
$^4$~Massachusetts Institute of Technology, Cambridge, MA 02139-4307, USA\\
$^5$~National Central University, Chung-Li, Taiwan\\
$^6$~University of Illinois at Chicago, Chicago, IL 60607-7059, USA\\
$^7$~University of Maryland, College Park, MD 20742, USA\\
$^8$~University of Rochester, Rochester, NY 14627, USA\\
}

\begin{abstract}
\noindent

We present results on two-particle angular correlations in proton-proton collisions 
at center of mass energies of 200 and 410~GeV. The PHOBOS experiment at the
Relativistic Heavy Ion Collider has a uniquely large coverage for charged particles, 
giving the opportunity to explore the correlations at both short- and long-range scales. 
At both energies, a complex two-dimensional correlation structure in $\Delta \eta$ and $\Delta \phi$ is observed.
In the context of an independent cluster model of short-range correlations, the cluster size 
and its decay width are extracted from the two-particle pseudorapidity correlation function 
and compared with previous measurements in proton-proton and proton-antiproton collisions, 
as well as PYTHIA and HIJING predictions. 

\vspace{3mm}
\end{abstract}

\maketitle

\section{Introduction}

Studies of multiparticle correlations have proven to be a powerful tool in exploring 
the underlying mechanism of particle production in high energy hadronic collisions. 
Inclusive two-particle correlations have been found (e.g. Ref.~\cite{UA5_3energy}) to 
have two components: direct two-particle correlations as well as an effective ``long-range'' 
correlation due to event-by-event fluctuations of the overall particle multiplicity.
By considering the two-particle rapidity 
density at fixed multiplicity, the intrinsic correlations between particles were isolated 
and found to be approximately Gaussian, with a range of $\sigma_{\eta} \sim 1$ unit 
in pseudorapidity ($\eta=-\ln(\tan(\theta/2))$). Thus, these correlations have been conventionally 
called ``short-range'' correlations. Their properties have been explained by the concept of ``cluster'' emission.

The simple idea that hadrons are produced in clusters, rather than individually, has had 
great success in describing many features of particle production \cite{UA5_3energy,ISR_twolowenergy}. 
In a scenario of independent cluster emission, clusters are formed before the final-state 
hadrons and are independently emitted according to a dynamically generated distribution in 
$\eta$ and $\phi$. The clusters subsequently decay isotropically in their own rest frame 
into the observed final-state hadrons. An independent cluster emission model has been 
widely applied to the study of two-particle correlations \cite{cluster_model,cluster_fit},
where the observed correlation strength and extent in phase space can be parameterized in 
terms of the cluster multiplicity, or ``size'' (the average number of particles in a cluster) 
and the decay ``width'' (the separation of the particles in pseudorapidity). However, it should be noted that independent cluster
emission is only a phenomenological approach which provides no insight about cluster production
mechanisms. Further modeling will be required to connect these studies to the underlying QCD dynamics. 

A related measurement, studying the forward-backward multiplicity correlations, was performed 
in Au+Au collisions at center of mass energy per nucleon pair ($\sqrt{s_{NN}}$) of 200~GeV 
using the PHOBOS detector at the Relativistic Heavy Ion Collider (RHIC) \cite{FB_corr}. 
The event-by-event observable $C=(N_{F}-N_{B})/\sqrt{N_{F}+N_{B}}$ is constructed, where $N_{F}$ 
and $N_{B}$ are defined to be the total multiplicity in two symmetric regions forward and backward 
of mid-rapidity. The variance ($\sigma_{C}^{2}$), which is related to the cluster size, was measured. 
An effective cluster size of approximately 2-3, increasing with the size of the pseudorapidity window, was observed.
In heavy ion collisions at RHIC, it has been predicted that the formation of a quark gluon plasma
(QGP) could modify cluster properties relative to p+p collisions \cite{AAcluster_prediction}.
Unfortunately, the method used in Ref.~\cite{FB_corr} was found to have intrinsic 
limitations for measuring the properties of clusters emitted near mid-rapidity.  
This makes direct tests of these predictions difficult and suggests a need for different 
methods to access cluster properties directly.

In order to explore the properties of short-range correlations in more detail and to study 
cluster properties quantitatively, this paper presents the two-particle angular correlations 
in p+p collisions at center of mass energies ($\sqrt{s}$) of 200 and 410~GeV over a very broad acceptance 
in $\Delta \eta$ ($ = \eta_{1} - \eta_{2}$) and $\Delta \phi$ ($ = \phi_{1} - \phi_{2}$).
The PHOBOS Octagon detector, covering pseudorapidity $-3<\eta<3$ over almost the full azimuth,   
is well suited to measure the correlations between particles emitted from clusters,
which have been found to extend up to $\Delta \eta$ = 2-3 units of pseudorapidity.
The observed two-dimensional (2-D) correlation function shows a complex correlation structure 
in the hadronic final state. In this analysis, both the cluster size and its decay
width are extracted from the two-particle correlations as a function of $\Delta \eta$.
Furthermore, the energy, multiplicity and $\Delta \phi$ dependence of the cluster size and 
decay width are studied and compared to previous measurements as well as PYTHIA \cite{PYTHIA} 
and HIJING \cite{HIJING} predictions at various energies.
Our present study in p+p collisions will provide a useful baseline measurement 
for understanding the hadronization stage in A+A collisions.

\section{DATA SETS}

The data presented here for p+p collisions at $\sqrt{s} =$ 200 and 410~GeV were collected 
during RHIC Run 4 (2004) and Run 5 (2005) using the large-acceptance PHOBOS Octagon multiplicity 
detector covering pseudorapidity $-3<\eta<3$ over almost full azimuth. A full description of 
the PHOBOS detector can be found in Ref.~\cite{phobos_detector}. Single diffractive p+p 
events were suppressed by requiring at least one hit in each of two sets of 16 scintillator
trigger counters located at distances of $-3.21$ m and $3.21$ m from the
nominal interaction point $z_{\rm vtx}=0$ along the beam axis.
They cover an acceptance of $3<|\eta|<4.5$ and $-180^{\circ}<\phi<180^{\circ}$. About 0.5 million 
200~GeV and 0.8 million 410~GeV p+p events were selected for further analysis by requiring that
the main collision vertex fell within $|z_{\rm vtx}|<10$~cm along the beam axis.

The angular coordinates ($\eta,\phi$) of charged particles are measured using the
energy deposited in the silicon pads of the Octagon. The granularity of the Octagon is determined 
by the sizes of the readout pads which are about $11.25^{\circ}$ ($\sim$ 0.2 radians) in $\phi$ and range 
from 0.006 to 0.05 in $\eta$. Monte-Carlo studies using the PYTHIA event generator 
show that, in the low-multiplicity p+p environment, 98\% of hits on adjacent pads 
in $\eta$ are created by single primary particles with a small angle of incidence. 
All such neighboring hits are merged to single hits in order to avoid fake particle pairs. 
Noise and background hits are rejected by cutting on the deposited energy 
corrected for the path length through the silicon after hit merging, assuming that the charged 
particle originated from the main vertex. Depending on $\eta$, the merged hits with 
less than 50-60\% of the energy loss expected for a minimum ionizing particle are rejected.
More details of the hit reconstruction procedure can be found in Ref.~\cite{hits}.

\section{Definition of the two-particle correlation function}

Following an approach similar to that in Ref.~\cite{ISR_twolowenergy}, the inclusive charged two-particle 
correlation function in two-particle ($\eta_{1},\eta_{2},\phi_{1},\phi_{2}$) space is defined as follows:

\vspace{-0.4cm}
\begin{equation}
\label{2pcorr_incl_4D}
R(\eta_{1},\eta_{2},\phi_{1},\phi_{2}) =
\left<(n-1)\left(\frac{\rho_{n}^{\rm II}(\eta_{1},\eta_{2},\phi_{1},\phi_{2})}
{\rho_{n}^{\rm I}(\eta_{1},\phi_{1})\rho_{n}^{\rm I}(\eta_{2},\phi_{2})}-1\right)\right>
\end{equation}

\noindent where, 
\[ \rho_{n}^{\rm I}(\eta,\phi)=\frac{1}{n\sigma_{n}}
\frac{d^{2}\sigma_{n}}{d\eta d\phi} \] 
is the single charged particle density distribution and 
\[ \rho_{n}^{\rm II}(\eta_{1},\eta_{2},\phi_{1},\phi_{2})=
\frac{1}{n(n-1)\sigma_{n}}\frac{d^{4}\sigma_{n}}{d\eta_{1} 
d\eta_{2} d\phi_{1} d\phi_{2}} \] 
denotes the charged particle pair distribution. Here $\sigma_{n}$ is the cross section of 
observing $n$ charged particles. The above distributions obey the normalization relations: 
\[ \int  \rho_{n}^{\rm I}(\eta,\phi) \,d\eta d\phi = 1\] and
\[ \int  \rho_{n}^{\rm II}(\eta_{1},\eta_{2},\phi_{1},\phi_{2}) 
\,d\eta_{1}d\eta_{2}d\phi_{1}d\phi_{2} = 1.\]
 
%\noindent The correlatioin function is constructed with fixed multiplicity first and then 
%\noindent Defined in the form of a ratio, any inefficiencies and acceptance effects of the 
%detector will cancel out.

In practice, we concentrate on the difference in azimuthal angle and pseudorapidity 
between two particles. The correlation function in Eq.~\ref{2pcorr_incl_4D} is 
simplified by averaging over the full acceptance $-3<\eta_{1},\eta_{2}<3$ and $-180^{\circ}<\phi_{1},\phi_{2}<180^{\circ}$,
reducing the dimensionality of the parameter space to $\Delta \eta$ and $\Delta \phi$ with a range of
$|\Delta \eta|<6$ and $|\Delta \phi|<180^{\circ}$:

\vspace{-0.4cm}
\begin{equation}
\label{2pcorr_incl}
R(\Delta \eta,\Delta \phi)=\left<(n-1)\left(\frac{\rho_{n}^{\rm II}
(\Delta \eta,\Delta \phi)}{\rho^{\rm mixed}(\Delta \eta,\Delta \phi)}-1\right)\right>.
\end{equation} 

\noindent The pair distribution $\rho_{n}^{\rm II}(\Delta \eta,\Delta \phi)$ (with unit integral) 
is determined by taking particle pairs from the same event, then averaging over all events.
The mixed-event background, $\rho^{\rm mixed}(\Delta \eta,\Delta \phi)$ (with unit integral), 
is constructed by randomly selecting single particles from two different events with similar  
vertex position, representing a product of two single particle distributions. The vertex bin 
size of 0.5~cm used in event-mixing is mainly determined by the vertex resolution of the detector 
in p+p collisions. Since the background is found to be multiplicity independent, we use the inclusive
$\rho^{\rm mixed}(\Delta \eta,\Delta \phi)$ in our calculations of Eq.~\ref{2pcorr_incl}. 
The track multiplicity $n$ is introduced to compensate for trivial dilution effects from uncorrelated 
particles \cite{ISR_twolowenergy}. This stems from the fact that the number of uncorrelated pairs 
grows quadratically with $n$, while the number of correlated pairs grows only linearly. 

\section{Corrections and Systematics}

\begin{figure*}[th!]
\hspace{1.5cm}
\centerline{
  \mbox{\includegraphics[width=\linewidth]{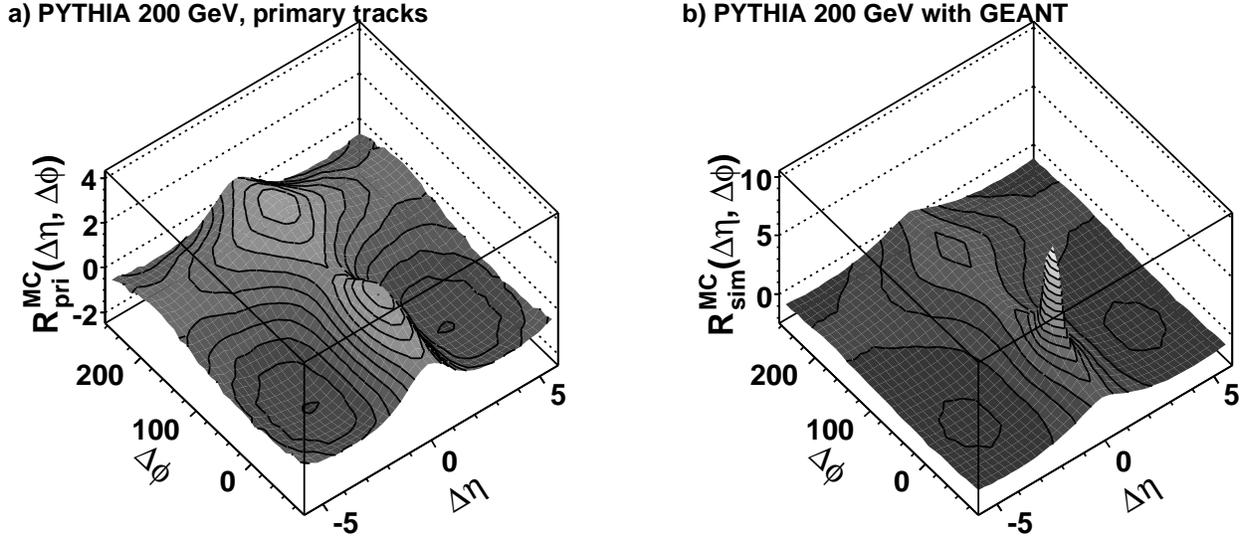}}}
\vspace{-0.3cm}
  \caption{ \label{corr_2D_rawpmc}
  Two-particle correlation function in $\Delta \eta$ and $\Delta \phi$ for a) generator level PYTHIA for 
  primary hadrons and b) PYTHIA with full GEANT detector 
  simulation and reconstruction procedures in p+p collisions at $\sqrt{s} =$ 200~GeV.}
\vspace{-0.3cm}
\end{figure*}

\begin{figure*}[thb!]
\vspace{-0.5cm}
\hspace{1.5cm}
\centerline{
  \mbox{\includegraphics[width=\linewidth]{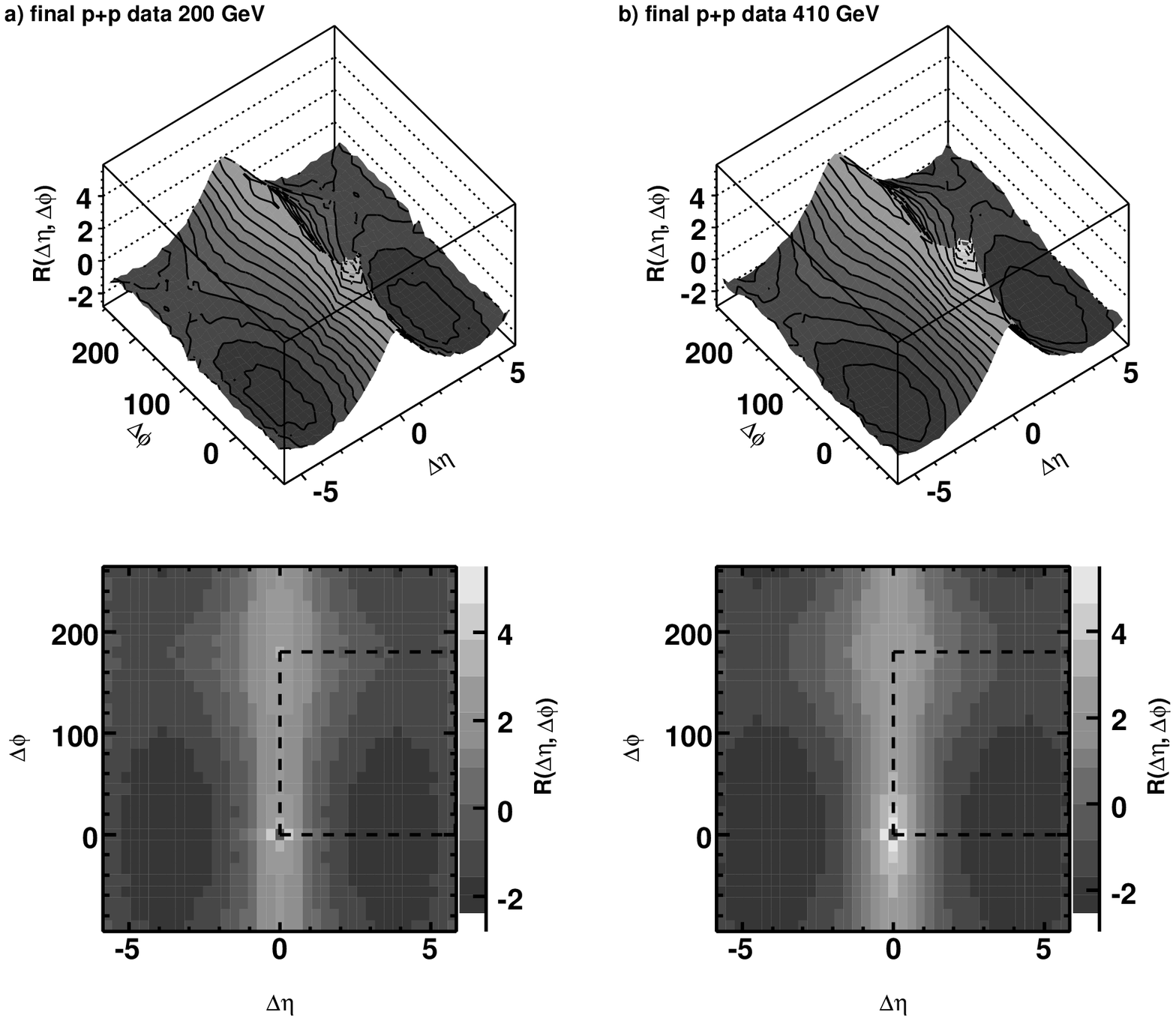}}}
\vspace{-0.3cm}
  \caption{ \label{corr_2D_finaldata_bothenergy}
  Two-particle angular correlation function in $\Delta \eta$ and $\Delta \phi$
  in p+p collisions at $\sqrt{s} =$ 200~GeV (column a) and 410~GeV (column b)
  (regions of $|\Delta \eta|<0.15$ and $|\Delta \phi|<5.625^{\circ}$ are excluded)
  shown in two types of representations. The areas enclosed by the dashed lines on the lower 
  two panels indicate the actual information in the data which is then reflected to 
  other regions to more clearly show the full shape of the correlation.}
\vspace{-0.3cm}
\end{figure*}

The PHOBOS Octagon detector has a single layer of silicon, and so does not provide momentum, 
charge or mass information for the observed particles. Thus secondary effects, such as 
$\delta$-electrons, $\gamma$ conversions and weak decays cannot be directly rejected.
These will contribute to correlations unrelated to those between primary hadrons, and 
modify the shape of the measured correlation function. The incomplete azimuthal
acceptance in some pseudorapidity regions naturally suppresses the overall correlation strength, 
but MC simulations show that it does not change the shape of the correlation in 
$\Delta \eta$ and $\Delta \phi$. To correct for these detector effects in the data,
correlation functions are calculated for PYTHIA events at $\sqrt{s}$ = 200~GeV
both at the generator level for true primary charged hadrons, 
$R_{\rm pri}^{\rm MC}(\Delta \eta, \Delta \phi)$ (Fig.~\ref{corr_2D_rawpmc}a) 
and with the full GEANT detector simulation and reconstruction procedure, 
$R_{\rm sim}^{\rm MC}(\Delta \eta, \Delta \phi)$ (Fig.~\ref{corr_2D_rawpmc}b). 
As will be illustrated in Fig.~\ref{corr_2D_finaldata_bothenergy}, 
the actual information in $R(\Delta \eta, \Delta \phi)$ is reflected to 
the full $\Delta \eta$ and $\Delta \phi$ range 
in order to more clearly show the shape of the correlation function.
The whole correction procedure can be summarized by the following equation:

\vspace{-0.4cm}
\begin{equation}
\label{corrected_corr_scale_corrected}
R_{\rm final}^{\rm data}(\Delta \eta, \Delta \phi) = A \times [R_{\rm raw}^{\rm data}
(\Delta \eta, \Delta \phi)-S(\Delta \eta, \Delta \phi)].
\end{equation}

\begin{figure*}[htb!]
%\hspace{1.0cm}
\centerline{
  \mbox{\includegraphics[width=\linewidth]{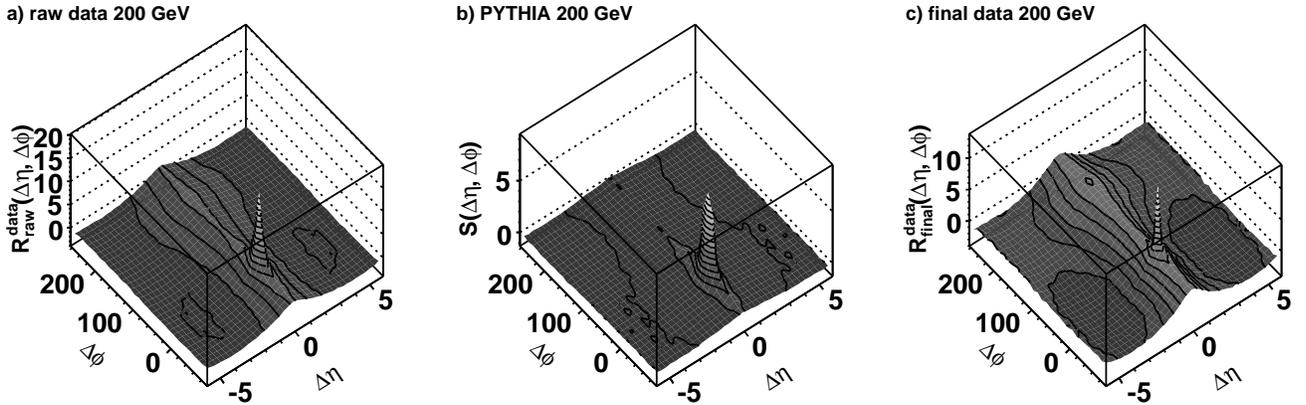}}}
\vspace{-0.3cm}
  \caption{ \label{corr_2D_rawdata}
            a) Raw two-particle correlation function for p+p collision data, 
            b) correction function $S(\Delta \eta, \Delta \phi)$ from PYTHIA
            and c) final two-particle correlation function for p+p collision data
            in $\Delta \eta$ and $\Delta \phi$ at $\sqrt{s} =$ 200~GeV.}
\vspace{-0.5cm}
\end{figure*}

The overall correlation structure consists of both intrinsic and secondary correlations 
and these two sources of correlations are found to be largely independent of each other in MC studies, 
i.e.\ the correlation from secondaries is mostly determined by sensor thickness, detector geometry, 
known cross-sections and decay kinematics. First, within a narrow vertex range (2cm), 
we compare the generator level MC correlation function excluding particles outside the
PHOBOS detector acceptance, $R_{\rm pri,acc}^{\rm MC}(\Delta \eta, \Delta \phi)$, 
to the correlation function observed after processing the same MC events with all the primary hadrons 
through the GEANT simulation, $R_{\rm sim}^{\rm MC}(\Delta \eta, \Delta \phi)$ (Fig.~\ref{corr_2D_rawpmc}b). 
The difference between the two correlation functions, $S(\Delta \eta, \Delta \phi)$
(Fig.~\ref{corr_2D_rawdata}b), is attributed to the effects of secondary 
interactions, weak decays, and the reconstruction procedure:

\vspace{-0.4cm}
\begin{equation}
\label{secondary_corr}  
S(\Delta \eta, \Delta \phi) =R_{\rm sim}^{\rm MC}(\Delta \eta, \Delta \phi)
-R_{\rm pri,acc}^{\rm MC}(\Delta \eta, \Delta \phi).
\end{equation}

To estimate the suppression of the correlation strength due to limited acceptance,
we compare $R_{\rm pri}^{\rm MC}(\Delta \eta, \Delta \phi)$ and
$R_{\rm pri,acc}^{\rm MC}(\Delta \eta, \Delta \phi)$ using 
a $\chi^{2}$ test to extract a scaling factor, $A$, for each vertex bin: 

\vspace{-0.4cm}
\begin{equation}
\label{acceptance_corr} 
\chi^{2} = \int \frac{[R_{\rm pri}^{\rm MC}(\Delta \eta, \Delta \phi)-A \times R_{\rm pri,acc}^{\rm MC}(\Delta \eta, \Delta \phi)]^{2}}
{\sigma_{\rm pri}^{\rm MC}(\Delta \eta, \Delta \phi) \times A \times \sigma_{\rm pri,acc}^{\rm MC}(\Delta \eta, \Delta \phi)} \, d\Delta \eta d\Delta \phi
\end{equation}

\noindent where $\sigma_{\rm pri}^{\rm MC}(\Delta \eta, \Delta \phi)$ and $\sigma_{\rm pri,acc}^{\rm MC}(\Delta \eta, \Delta \phi)$ are
the uncertainties of the correlation function $R_{\rm pri}^{\rm MC}(\Delta \eta, \Delta \phi)$ and $R_{\rm pri,acc}^{\rm MC}(\Delta \eta, \Delta \phi)$.
This scaling factor, $A$, is independent of $\Delta \eta$ and $\Delta \phi$,
and is found to vary between 1.3 and 1.5 over the vertex range used in this analysis.

To get the final correlation function, $R_{\rm final}^{\rm data}(\Delta \eta, \Delta \phi)$,
we apply Eq.~\ref{corrected_corr_scale_corrected} to the raw correlation
function, $R_{\rm raw}^{\rm data}(\Delta \eta, \Delta \phi)$. The correction term $S(\Delta \eta, \Delta \phi)$
from Eq.~\ref{secondary_corr} is used to subtract the effects
of secondaries and weak decays, and the scaling factor, $A$, from the $\chi^{2}$ test in Eq.~\ref{acceptance_corr} is used
to remove the suppression due to the holes in the PHOBOS acceptance.
This procedure is done separately for each vertex bin.
Three different MC generators are used to estimate the systematic
uncertainties to the correlation function from this correction procedure, including PYTHIA,
HIJING and a modified PYTHIA in which all intrinsic correlations have been removed
by performing event mixing at the primary hadron level (i.e.\ before weak decays). Typically the systematic
error (biggest for the peak at $\Delta \eta$=0 and $\Delta \phi$=0) is less than 5\%.

Both the correlation function for the data (Fig.~\ref{corr_2D_rawdata}a) and for the PYTHIA simulations (Fig.~\ref{corr_2D_rawpmc}b)
have a sharp peak at small $\Delta \eta$ and $\Delta \phi$, which is twice as high in the data as in the fully simulated MC events,
but which is not present in the analysis using primary particles from the MC generator (Fig.~\ref{corr_2D_rawpmc}a). In the full GEANT simulations, 
the particles contributing to this peak are found to be mainly $\delta$-electrons (50\%) and 
$\gamma$ conversions (40\%). The width of the peak is about 0.3 in $\Delta \eta$ and $28^{\circ}$ 
in $\Delta \phi$ for both data and MC, indicating the same origin. 
However, the final data still contain a much narrower peak at the near-side of 
$R_{\rm final}^{\rm data}(\Delta \eta, \Delta \phi)$ (Fig.~\ref{corr_2D_rawdata}c). It is 
likely that this small angle structure results from background and detector effects which 
are not included in the MC simulation, although it is not possible to rule out unknown 
physics effects not implemented in the event generators.  Since the physics of the cluster-like 
particle production investigated in this analysis is dominated by correlations on scales of 
approximately one unit in $\Delta \eta$, as will be shown later, we proceed by rejecting 
pairs in a small two-particle acceptance of $|\Delta \eta|<0.15$ and $|\Delta \phi|<5.625^{\circ}$ 
(the single bin centered at $\Delta \eta$=0 and $\Delta \phi$=0). Studies using primary particles 
from MC generators and the fully simulated events show that the extracted cluster parameters, 
described in the next section, change by less than 0.1\% due to this cut. 

In addition to the systematic errors related to the correction procedure, other systematic uncertainties 
are calculated by varying the vertex position and hit 
threshold cuts, and by studying the time dependence of the results within the two data sets. 
Since the acceptance of our detector is strongly dependent on vertex position, any systematic uncertainties due to 
acceptance and geometrical description should manifest themselves as a dependence of the 
final results on vertex position. The hit threshold cuts are used to discriminate between 
primary particles and noise or background hits; a variation of the thresholds allows us 
to test the impact of noise and background hits on the final results. For each selection of reconstruction 
parameters, a set of correlation functions are constructed under different conditions 
(i.e.\ 10 different vertex positions), and fully corrected using the MC procedure described above.
The RMS of these correlation functions is calculated as an estimate of the systematic error from each particular 
source. The systematic errors due to the vertex position dependency, hit threshold cuts and variations 
in different subsets of events are added in quadrature to get the total RMS for each bin 
in $\Delta \eta$ and $\Delta \phi$. The vertex position dependency turns out to be the dominant source of the uncertainties.
The final systematic uncertainties are quoted as 90\% C.L.(1.6 $\times$ RMS).

\section{Results}

The final two-particle inclusive correlation functions, averaged over 10 vertex bins,
are shown in Fig.~\ref{corr_2D_finaldata_bothenergy} as a function 
of $\Delta \eta$ and $\Delta \phi$ at $\sqrt{s}$ = 200 GeV (column a) and 410~GeV (column b). The near-side 
hole corresponds to the excluded region of $|\Delta \eta|<0.15$ and $|\Delta \phi|<5.625^{\circ}$. 
The systematic uncertainties in the absolute value of R($\Delta \eta$,$\Delta \phi$) are of the order of 0.3, 
relative to a peak value of 5, with little $\Delta \eta$ or $\Delta \phi$ dependence.

The complex 2-D correlation structure shown in Fig.~\ref{corr_2D_finaldata_bothenergy} is approximately 
Gaussian in $\Delta \eta$ and persists over the full $\Delta \phi$ range, becoming broader toward 
larger $\Delta \phi$ (which will be discussed in quantitative detail below).
Similar structures also exist in PYTHIA (Fig.~\ref{corr_2D_rawpmc}a) though they do
not reproduce the strength of the short-range rapidity correlations seen in the data.
The qualitative features of the observed correlation structure are consistent with
an independent cluster approach according to a simulation study from the ISR experiment
using a low-mass resonance ($\rho$,$\omega$,$\eta$) gas model \cite{ISR_twolowenergy}. 
The excess of the near-side peak ($\Delta \eta \sim 0$ and $\Delta \phi \sim 0$)
relative to the away-side could be partially a result of the HBT effect \cite{HBT}. 
This possibility is investigated in the Appendix using a simple MC model and found 
to be negligible for the cluster properties investigated below.

To study the correlation structure quantitatively, the 2-D correlation function is projected 
into a 1-D pseudorapidity correlation function of $\Delta \eta$ by integrating 
$\rho_{n}^{\rm II}(\Delta \eta,\Delta \phi)$ and $\rho^{\rm mixed}(\Delta \eta,\Delta \phi)$ over $\Delta \phi$ as follows:

\vspace{-0.4cm}
\begin{equation}
\label{2pcorr_1Dprojection}
R(\Delta \eta)=\left<(n-1)\left(\frac{\int \rho_{n}^{\rm II}
(\Delta \eta, \Delta \phi) d\Delta \phi}{\int \rho^{\rm mixed}(\Delta \eta, \Delta \phi) d\Delta \phi}-1\right)\right>.
\end{equation} 

\begin{figure}[bht!]
\hspace{-0.1cm}
\centerline{
  \mbox{\includegraphics[width=1.07\linewidth]{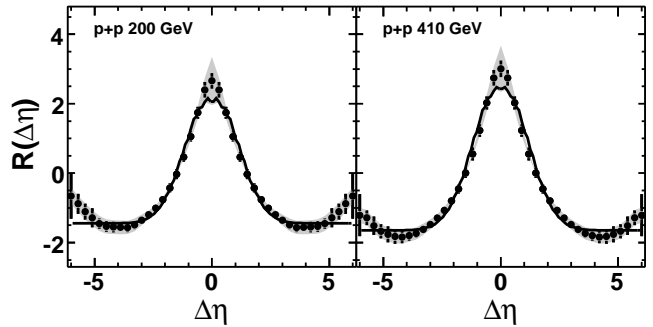}}}
\vspace{-0.3cm}
  \caption{ \label{2pcorr_eta}
  Two-particle pseudorapidity correlation function, averaged over the $\Delta \phi$ 
  range from $0^{\circ}$ to $180^{\circ}$, in p+p collisions 
  at $\sqrt{s} =$ 200~GeV (left) and 410~GeV (right). 
  The solid curves correspond to the fits by the cluster model
  using Eq.~\ref{2pcorr_clusterfitting_incl} over the full $\Delta \eta$ range.
  The error bars and bands correspond to point-to-point systematic errors 
  and overall scale errors respectively with 90\% C.L. The statistical errors are
  negligible.}
\vspace{-0.3cm}
\hspace{-0.3cm}
\end{figure}

\begin{figure}[ht]
\hspace{-0.3cm}
\centerline{
  \mbox{\includegraphics[width=0.95\linewidth]{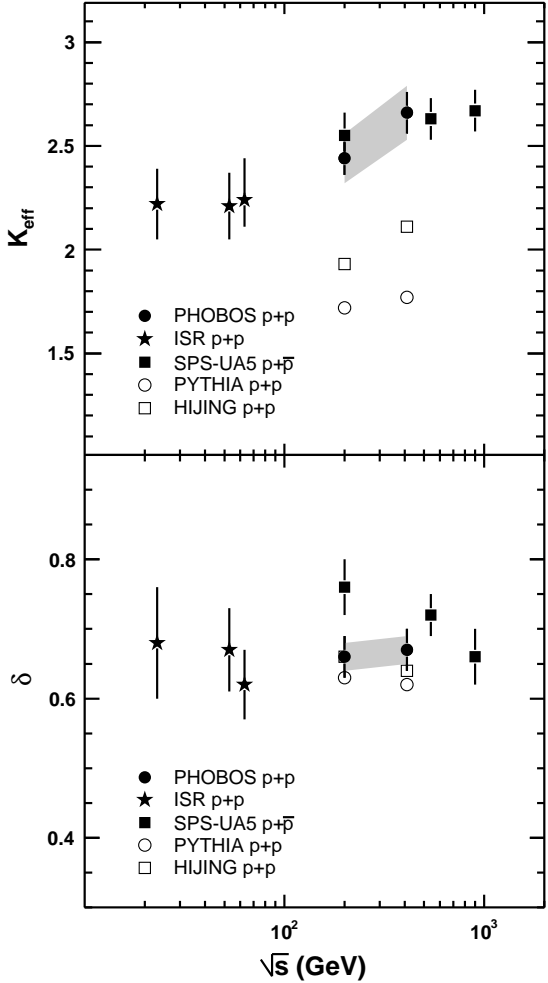}}}
\vspace{-0.3cm}
  \caption{$K_{\rm eff}$ (top) and $\delta$ (bottom) as a function of $\sqrt{s}$ 
            measured by PHOBOS in solid circles, as well as UA5 \cite{UA5_3energy}
            (solid squares) and ISR \cite{ISR_twolowenergy, ISR_63GeV}(solid stars) 
            experiments for p+p and p+\={p} collisions. Open circles and squares 
            show the PYTHIA and HIJING results respectively. The error representations 
            are identical to those in Fig.~\ref{2pcorr_eta}. }
  \label{cluster_sqrts}
\vspace{-0.3cm}
\end{figure}

\begin{figure}[t!]
\hspace{-0.3cm}
\centerline{
  \mbox{\includegraphics[width=0.95\linewidth]{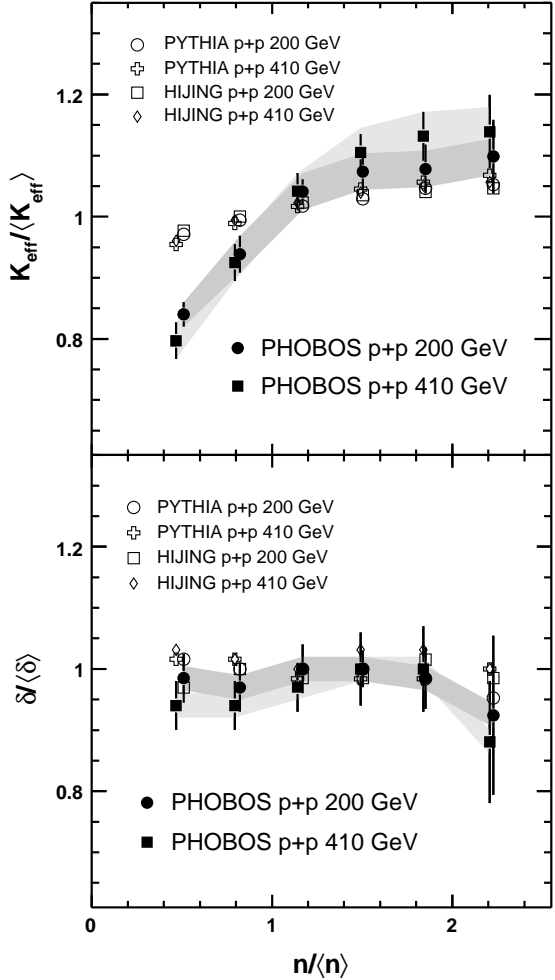}}}
\vspace{-0.3cm}
  \caption{Normalized effective cluster size $K_{\rm eff}/\langle K_{\rm eff} \rangle$ 
           (top) and decay width $\delta/\langle \delta \rangle$ (bottom) as a function 
           of normalized multiplicity $n/\langle n \rangle$ in p+p collisions at 
           $\sqrt{s} =$ 200 and 410~GeV measured by PHOBOS (solid symbols), 
           as well as MC studies (open symbols).  The error representations 
           are identical to those in Fig.~\ref{2pcorr_eta}.}
  \label{cluster_n}
\vspace{-0.3cm}
\end{figure}

\begin{figure}[t!]
\hspace{-0.3cm}
\centerline{
  \mbox{\includegraphics[width=0.95\linewidth]{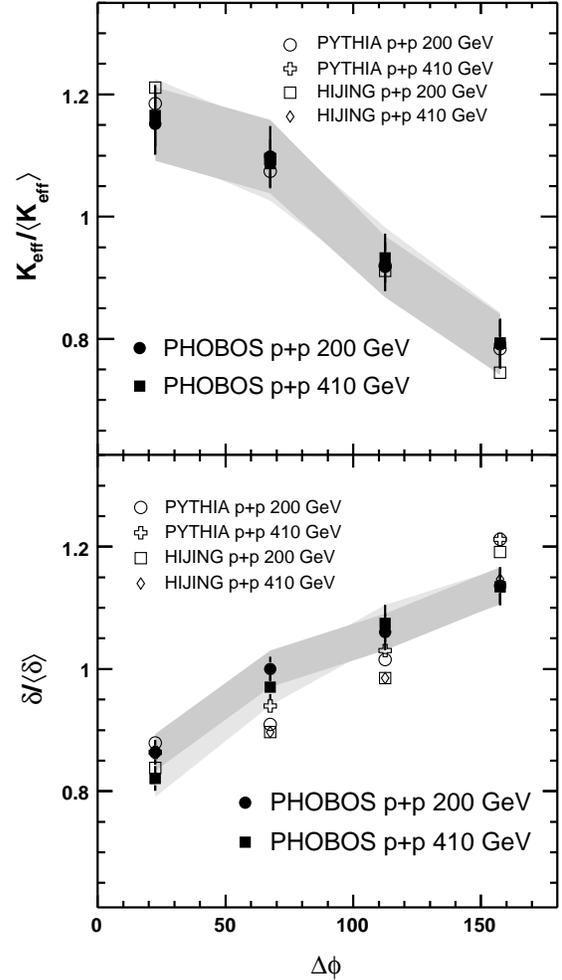}}}
\vspace{-0.3cm}
  \caption{Normalized effective cluster size $K_{\rm eff}/\langle K_{\rm eff} \rangle$ 
           (top) and decay width $\delta/\langle \delta \rangle$ (bottom) as a function 
           of $\Delta \phi$ in p+p collisions at $\sqrt{s}$ = 200 and 410~GeV measured by 
           PHOBOS (solid symbols), as well as MC studies (open symbols). The error representations 
           are identical to those in Fig.~\ref{2pcorr_eta}.}
  \label{cluster_phi}
\vspace{-0.3cm}
\end{figure}

\begin{figure}[h!]
\hspace{-0.1cm}
\centerline{
  \mbox{\includegraphics[width=1.07\linewidth]{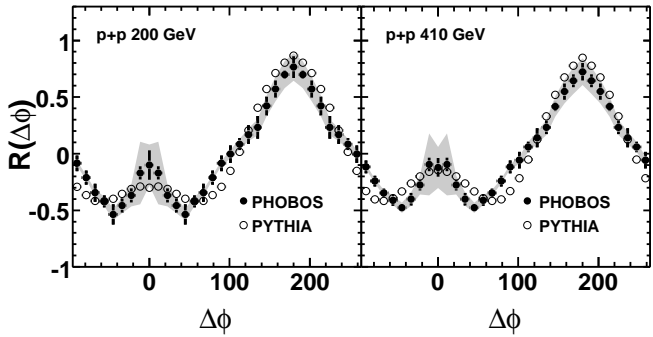}}}
\vspace{-0.3cm}
  \caption{ \label{2pcorr_phi}
  Two-particle azimuthal correlation function, averaged over the $\Delta \eta$ 
  range from 0 to 6, in p+p collisions at $\sqrt{s} =$ 200~GeV (left) and 410~GeV (right) 
  from PHOBOS (solid circles), as well as MC studies (open circles).
  The error representations are identical to those in Fig.~\ref{2pcorr_eta}.}
\vspace{-0.3cm}
\end{figure}

\noindent The two-particle pseudorapidity correlation function R($\Delta \eta$), averaged over the $\Delta \phi$ 
range from $0^{\circ}$ to $180^{\circ}$, is shown in Fig.~\ref{2pcorr_eta} at $\sqrt{s} =$ 200 and 410~GeV.
The error bars (also in Figs.~\ref{cluster_sqrts} -~\ref{2pcorr_phi}) 
correspond to point-to-point systematic errors with 90\% C.L.
The error bands (also in Figs.~\ref{cluster_sqrts} -~\ref{2pcorr_phi}) 
denote an overall scale error with 90\% C.L. as an indication of the uncertainties in the correction method 
which tends to move all of the data points up and down in a correlated fashion. The statistical errors are
negligible due to the large p+p event sample used in this analysis.

In the context of an independent cluster emission model, $R(\Delta \eta)$ takes the functional form \cite{cluster_fit}:

\vspace{-0.4cm}
\begin{equation}
\label{2pcorr_clusterfitting_incl}
R(\Delta \eta)=\alpha\left[\frac{\Gamma(\Delta \eta)}{\rho^{\rm mixed}(\Delta \eta)}-1\right]  
\end{equation}

\noindent where the correlation strength $\alpha=\frac{ \langle K(K-1) \rangle }{ \langle K \rangle }$ is a parameter containing information on the 
distribution of cluster size $K$. The function $\Gamma(\Delta \eta)$ is a Gaussian function \[ \propto exp{[- (\Delta \eta)^{2}/(4\delta^{2})]} \] 
characterizing the correlation of particles originating from a single cluster where $\delta$ indicates 
the decay width of the clusters. The background distribution $\rho^{\rm mixed}(\Delta \eta)$ is just the distribution 
obtained by event-mixing introduced in section III. To correct for the holes in the PHOBOS acceptance, we calculate the 
ratio of the background for PYTHIA primary particles, $\rho_{\rm MC,pri}^{\rm mixed}(\Delta \eta)$, to the one obtained in the full 
GEANT simulations, $\rho_{\rm MC,sim}^{\rm mixed}(\Delta \eta)$. The ratio is applied to the background calculated from the data,
$\rho_{\rm data,raw}^{\rm mixed}(\Delta \eta)$, as a multiplicative factor:

\vspace{-0.4cm}
\begin{equation}
\label{background_corr}
\rho_{\rm data,final}^{\rm mixed}(\Delta \eta)=\frac{\rho_{\rm MC,pri}^{\rm mixed}(\Delta \eta)}
{\rho_{\rm MC,sim}^{\rm mixed}(\Delta \eta)} \times \rho_{\rm data,raw}^{\rm mixed}(\Delta \eta).
\end{equation}

The effective cluster size is related to the extracted correlation strength via the relation: 

\vspace{-0.4cm}
\begin{equation}
\label{Keff}
K_{\rm eff}=\alpha+1=\frac{\left<K(K-1)\right>}{\left<K\right>}+1=\left<K\right>+\frac{\sigma_{K}^{2}}{\left<K\right>}. 
\end{equation}

\noindent Without any knowledge of the distribution of $K$, it is impossible to directly 
measure the average cluster size $\langle K \rangle$. However, by a $\chi^{2}$ fit of Eq.~\ref{2pcorr_clusterfitting_incl} 
to the measured two-particle pseudorapidity correlation function, the effective cluster size $K_{\rm eff}$ 
and decay width $\delta$ can be estimated. The independent cluster model provides a good fit to the data 
over a large range in $\Delta \eta$, as shown in Fig.~\ref{2pcorr_eta}. An effective cluster size $K_{\rm eff}=2.44 \pm 0.08$ and width
$\delta = 0.66 \pm 0.03$ for $\sqrt{s}$ = 200~GeV and $K_{\rm eff}=2.66 \pm 0.10$, $\delta = 0.67 \pm 0.03$ 
for $\sqrt{s}$ = 410~GeV are obtained with scale errors of 5\% for $K_{\rm eff}$ and 3\% for $\delta$. The three most central 
points always lie above the fits, which could be due to residual secondary effects which are 
left uncorrected or other physics at this small scale in $\Delta \eta$. All fit results include these three points, 
but excluding them from the fit affects $K_{\rm eff}$ and $\delta$ by no more than 3\%. 

In Fig.~\ref{cluster_sqrts}, our data are compared with previous measurements of $K_{\rm eff}$ 
and $\delta$ as a function of $\sqrt{s}$.
At lower ISR energies \cite{ISR_twolowenergy,ISR_63GeV}, $K_{\rm eff}$ is constant within error bars. 
At the higher SPS energies \cite{UA5_3energy}, UA5 finds $K_{\rm eff}$ to be larger than at the ISR, 
but with little energy dependence between 200~GeV and 900~GeV. The PHOBOS data are in good agreement with 
the UA5 measurements and, with much higher statistics in the p+p event sample, show a clear energy dependence of $K_{\rm eff}$. 
By contrast, the cluster decay width $\delta$ remains almost constant over the full range of collision 
energies. The event generators, HIJING and PYTHIA, show a similar energy dependence of $K_{\rm eff}$ and $\delta$ 
to the data, but with a significantly lower magnitude of $K_{\rm eff}$.

The observed cluster size cannot be fully explained by a resonance decay model even at very low energies, 
since the expectation of $\langle K \rangle$ from resonance decays is about 1.5 (extrapolating to 1.7 for $K_{\rm eff}$ 
depending on the assumed cluster size distribution \cite{UA5_3energy}). This is significantly lower than 
the observed values, but is close to what is seen in PYTHIA. The HBT effect, after averaging over $\Delta \phi$, 
would increase the cluster size by no more than 2\% (see Appendix). Additional sources of short-range 
correlations, such as local quantum number conservation \cite{trainor_pp200}, are needed to describe the data.
As the energy increases, the onset of jets should play a more important role in the particle production giving 
bigger clusters, which could be the underlying cause for the observed energy dependence of $K_{\rm eff}$. 
At the LHC, with p+p collision at $\sqrt{s}$ = 14~TeV, jet-like particle production is expected to 
be dominant and should manifest itself in a further increase in the effective cluster size.

To gain further detailed information, the normalized cluster parameters, $K_{\rm eff}/\langle K_{\rm eff} \rangle$ 
and $\delta/\langle \delta \rangle$, are calculated as a function of the normalized charged multiplicity $n/\langle n \rangle$ at 
$\sqrt{s}$ = 200 and 410~GeV (Fig.~\ref{cluster_n}). Scaled by the average charged multiplicity $\langle n \rangle$, the distribution of 
$n/\langle n \rangle$ turns out to be essentially identical for data and charged primary tracks from MC (PYTHIA and HIJING),
despite the holes in the PHOBOS Octagon detector.
By dividing the cluster parameters at different multiplicity by the averaged values, a scaling behavior is observed
between two different energies in both data and MC. $K_{\rm eff}/\langle K_{\rm eff} \rangle$ increases with 
event multiplicity while $\delta/\langle \delta \rangle$ is found to be largely independent of it. 
PYTHIA and HIJING give a similar multiplicity dependence, but the increase in $K_{\rm eff}/\langle K_{\rm eff} \rangle$ 
is not as strong as in the data. Measurements from the ISR and UA5 experiments \cite{UA5_3energy, ISR_twolowenergy, SFM_2pcorr, ISR_63GeV} are qualitatively consistent with PHOBOS, but significantly limited by statistics, and are thus not shown here.

To explicitly show the $\Delta \phi$ dependence of the short-range pseudorapidity correlation seen in 
Fig.~\ref{corr_2D_finaldata_bothenergy}, the $\Delta \phi$ range from $0^{\circ}$ to $180^{\circ}$ is divided into four regions and projected 
separately onto the $\Delta \eta$ axis. Fig.~\ref{cluster_phi} shows $K_{\rm eff}/\langle K_{\rm eff} \rangle$ and $\delta/\langle \delta \rangle$
for different $\Delta \phi$ regions. $K_{\rm eff}/\langle K_{\rm eff} \rangle$ gradually decreases while $\delta$ increases 
as one goes from small to large $\Delta \phi$ region. This might reflect some information about the transverse momentum 
distribution of the clusters \cite{{ISR_twolowenergy}}. High $p_{T}$ clusters should generally contribute to a narrow hump 
in the near-side (near $\Delta \phi = 0^{\circ}$) of the correlation function in Fig.~\ref{corr_2D_finaldata_bothenergy}, 
whereas the broader away-side (near $\Delta \phi = 180^{\circ}$) comes from clusters with lower transverse 
momentum. Again, PYTHIA and HIJING are qualitatively similar to data but show a smaller decay width 
at intermediate $\Delta \phi$ and a larger decay width near $\Delta \phi = 180^{\circ}$.

The two-particle azimuthal correlation functions R($\Delta \phi$), averaged over a broad range of $\Delta \eta$
from 0 to 6, in p+p collisions at $\sqrt{s}$ = 200 and 410~GeV are presented in Fig.~\ref{2pcorr_phi}. 
R($\Delta \phi$) is obtained using a procedure similar to that shown
for R($\Delta \eta$) in Eq.~\ref{2pcorr_1Dprojection}. The observed asymmetric structure in $\Delta \phi$ 
(with the $\Delta\eta$-averaged away-side peak larger than the near-side peak)
could also provide some information about the momentum and size distribution of clusters. 
The $\Delta \eta$ integrated correlation function is similar in magnitude both for data and PYTHIA, 
despite the significant difference in the extracted $K_{\rm eff}$. More detailed 
modeling of the cluster properties is needed to fully explain many aspects of the complex two-particle correlation function.

\section{Conclusion}

PHOBOS has measured two-particle angular correlations over a wide range in $\Delta \eta$
and $\Delta \phi$ in p+p collisions at $\sqrt{s}$ = 200 and 410~GeV.
Short-range correlations are observed over the full range in $\Delta \phi$, with a maximum at $\Delta \eta = 0$
which becomes wider at larger $\Delta \phi$. In the context of the cluster 
model, the effective cluster size and decay width are extracted from the two-particle pseudorapidity correlation 
function, and compared with previous experiments, as well as HIJING and PYTHIA event generators. Dependence of the cluster size on 
both beam energy and scaled multiplicity is observed, while the cluster width is essentially constant. The short-range correlation 
strength (or equivalently the effective cluster size, $K_{\rm eff}$) exceeds the expectation from the decays of resonance particles, 
suggesting the need for other sources of short-range correlations. As mentioned in the introduction, 
cluster properties could be modified in A+A collisions relative to p+p collisions by the presence of a QGP\cite{AAcluster_prediction}. 
Future studies should clarify the evolution of cluster parameters from p+p to d+Au and A+A collisions at RHIC energies. 

% Phobos acknowledgements and funding credits
%
% Last edited 7-Sep-2004 by George Stephans
%
% For expanded acknowledgements, you can uncomment some or all of the following
%
We acknowledge the generous support of the Collider-Accelerator Department.
% (including RHIC project personnel) and Chemistry Departments at BNL.  We
% thank Fermilab and CERN for help in silicon detector assembly.  We thank the
% MIT School of Science and LNS for financial support.  
%
This work was partially supported by U.S. DOE grants 
DE-AC02-98CH10886,
DE-FG02-93ER40802, 
DE-FC02-94ER40818,  % MIT
DE-FG02-94ER40865, 
DE-FG02-99ER41099, and
W-31-109-ENG-38, by U.S. 
NSF grants 9603486, % Phobos TOF 
0072204,            % Rochester until 6/03
and 0245011,        % Rochester starting 6/03
by Polish KBN grant 1-P03B-062-27(2004-2007),
by NSC of Taiwan Contract NSC 89-2112-M-008-024, and
by Hungarian OTKA grant (F 049823).

\section*{Appendix: HBT Correlations}
The Bose-Einstein or HBT correlation will certainly contribute to the two-particle angular 
correlation function and thus might impact the estimation of cluster parameters \cite{HBT}. 
It is known to cause a strong correlation between two identical particles 
having small relative invariant four-momentum $\textbf{q}_{inv}^{2}$ (=$(\textbf{q}_{1}-\textbf{q}_{2})^{2}$).
For identical bosons, the overall shape of the HBT effect can be approximated by a 1-D Gaussian correlation function:

\vspace{-0.3cm}
\begin{equation}
\label{hbt_cf}
C(\textbf{q}_{inv})=1+\lambda e^{-\textbf{q}_{inv}^{2}R_{inv}^{2}}.
\end{equation}
%\vspace{0.2cm}

To estimate the effect of the HBT correlation, in PYTHIA, which does not include this effect, we calculate 
the $\textbf{q}_{inv}$ for each pair in $\rho_{n}^{\rm II}(\Delta \eta,\Delta \phi)$ (ignoring the particle species), 
and weight the pair by C($\textbf{q}_{inv}$) to artificially 
introduce the HBT correlation. In our study, we use $\lambda$ = 0.8 and $R_{inv}$ = 1.0 fm. 
Considering the fact that most of the pairs measured in this analysis are not identical-particle pairs, which should
reduce the magnitude of $\lambda$, this choice of parameters exaggerates the expected effect of the HBT correlation.
Thus, it gives a conservative estimate of its potential effect on the data. 

\begin{figure}[htb]
\hspace{-0.3cm}
\centerline{
   \mbox{\includegraphics[width=0.85\linewidth]{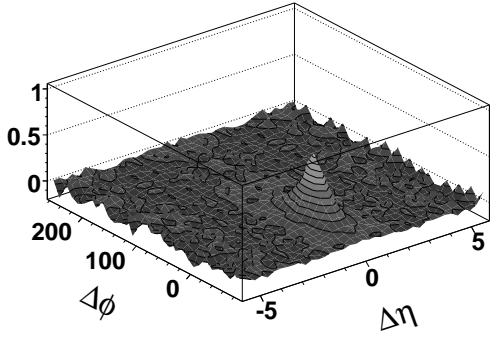}}}
\vspace{-0.2cm}
\caption{Difference of the correlation function with HBT weighting minus the 
         one without HBT weighting in $\Delta \eta$ and $\Delta \phi$ for PYTHIA at $\sqrt{s}$ = 200~GeV}
\label{corrdiff_2D_pythia200}
\end{figure}

\begin{figure}[htb]
\hspace{-0.3cm}
\centerline{
   \mbox{\includegraphics[width=0.85\linewidth]{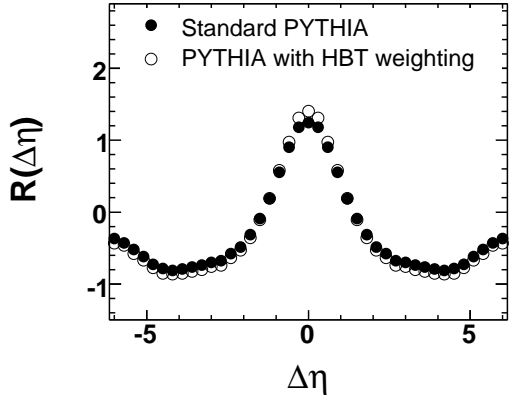}}}
\vspace{-0.2cm}
\caption{Comparison of the pseudorapidity correlation function with (open circles) and 
         without (solid circles) HBT weighting for PYTHIA at $\sqrt{s}$ = 200~GeV. }
\label{corrcomp_eta_pythia200}
\end{figure}

In Fig.~\ref{corrdiff_2D_pythia200}, the difference between the correlation function with and without 
C($\textbf{q}_{inv}$) weighting is shown as a function of $\Delta \eta$ and $\Delta \phi$ 
for PYTHIA. Finally, Fig.~\ref{corrcomp_eta_pythia200} shows a comparison of pseudorapidity correlation 
functions with and without HBT weighting, with extracted cluster sizes of 1.76 and 1.72 respectively.
From these studies, it appears that the HBT effect does enhance the short-range correlations 
with a range of around one unit in $\Delta \eta$ and $45^{\circ}$ in $\Delta \phi$. However, after averaging over 
$\Delta \phi$, it only slightly influences the rapidity correlation function, resulting in an 
increase in cluster size of at most 2\%. Even in the region $0^{\circ}<\Delta \phi<45^{\circ}$,
the increase is less than 3.5\%.

\end{document}